# Direction of Arrival Estimation of Sound Sources Using Icosahedral CNNs

David Diaz-Guerra, *Student Member, IEEE,* Antonio Miguel and Jose R. Beltran

*Abstract*—In this paper, we present a new model for Direction of Arrival (DOA) estimation of sound sources based on an Icosahedral Convolutional Neural Network (CNN) applied over SRP-PHAT power maps computed from the signals received by a microphone array. This icosahedral CNN is equivariant to the 60 rotational symmetries of the icosahedron, which represent a good approximation of the continuous space of spherical rotations, and can be implemented using standard 2D convolutional layers, having a lower computational cost than most of the spherical CNNs. In addition, instead of using fully connected layers after the icosahedral convolutions, we propose a new soft-argmax function that can be seen as a differentiable version of the argmax function and allows us to solve the DOA estimation as a regression problem interpreting the output of the convolutional layers as a probability distribution. We prove that using models that fit the equivariances of the problem allows us to outperform other state-of-the-art models with a lower computational cost and more robustness, obtaining root mean square localization errors lower than $10°$ even in scenarios with a reverberation time $T_{60}$ of 1.5 s.

## I. Introduction

DIRECTION of Arrival (DOA) estimation and Sound Source Localization (SSL) and tracking are topics that have been studied for decades by the signal processing community, but a technique with good performance in reverberant scenarios and low computational cost is yet to be found. Although it was originally studied using classical signal processing methods, such as the Generalized Cross-Correlation (GCC) functions [1] or the Multiple Signal Classification (MUSIC) [2], [3] and the Steered Response Power with Phase Transform (SRP-PHAT) [4], [5] algorithms, in the recent years the focus has turned towards deep learning techniques.

When we are designing a DOA estimation system based on neural networks, we need to decide how we are going to train it and what we are going to use as:

1) *Input:* The most common input features are the inter-microphone GCCs [6], [7], [8], and different spectral representations of the signals from each microphone [9], [10], [11], but other approaches have also been proposed, such as using Ambisonics intensity vectors [12], [13] or even the raw audio samples [14], [15] and, of course, combinations of several of the mentioned features [16], [17], [18].

2) *Network architecture:* Although Multi-layer Perceptrons (MLPs) were originally the most popular approach [6], [7], most current works propose the use of Convolutional Neural Networks (CNNs) [10], [19], [11], [17], sometimes including bi-directional recurrent layers at their end [9], [12], [18]. Modifying a CNN to be casual, i.e. making its Receptive Field (RF) include only past input features, is trivial, but the use of bi-directional recurrent layers makes the models unfeasible for real-time implementations since the backward direction is, by its own definition, completely anti-causal.

3) *Output:* Most of the first proposals performed the DOA estimation as a classification problem, so the output of the networks was a discretization of the search space whose maximum was considered to be the DOA estimation [10], [20], [21]. An important drawback of this approach is that the accuracy of the estimation is limited by the resolution of that discretization and, therefore, the size of the model grows too fast when trying to obtain both the azimuth and the elevation of the DOA or a 3D localization. To overcome this limitation, many new proposals solve the DOA estimation as a regression problem, so the outputs of the networks are directly the coordinates of the DOA (or a normalized version of them). Although directly inferring the elevation and the azimuth might seem to be the most direct and logical approach, it has been proven that it is easier to train the networks to infer the 3D coordinates of the unit vector pointing towards the DOA [22], [23] and this is the most popular approach nowadays [9], [24], [18].

An extensive survey on deep learning-based SSL and DOA estimation methods can be found in [25].

One of the problems shared by most of the previous proposals is that the decisions about the inputs features and the network architecture seem to have been made independently, leading to the use of architectures that are not optimal for their inputs or that, at least, do not fit with their properties. This does not follow the ideas of Geometric Deep Learning [26], an approach that is growing popular in the deep-learning community in recent times and that advocates that network architectures should be able to exploit the symmetries of their inputs and the problem they are intended to solve.

For example, when we use a spectrogram as if it was a 2D image and perform a convolution operation over it, we have equivariance to both time and frequency shifts, which

This work was supported in part by the Regional Government of Aragon (Spain) with a grant for postgraduate research contracts (2017-2021) co-founded by the Operative Program FSE Aragon 2014-2020.

D. Diaz-Guerra, A. Miguel and J.R. Beltran are with the Department of Electronic Engineering and Communications, University of Zaragoza, Zaragoza 50018, Spain. (e-mail: ddga@unizar.es; amiguel@unizar.es; jrbelbla@unizar.es).

means that if we apply a temporal or frequency shift in the input, we will obtain the same output but with that same shift. This is interesting in the case of temporal shifts since the same pattern in a different time instant usually represents the same information, but it is not so positive in the case of frequency shifts, since the phase differences for a given DOA are frequency dependent.

In [27], in order to better exploit the equivariance of CNNs, we proposed the use of 2D equiangular SRP-PHAT maps as inputs of a 3D CNN (the third dimension was the time) but this approach still had two main drawbacks: i) equiangular grids oversample the poles and are not the optimal way of sampling a sphere and ii) the DOA estimation problem and the SRP-PHAT maps computed from compact arrays are equivariant to spherical rotations, not to Cartesian translations on those equiangular projections.

Many new kinds of CNNs have been proposed in recent years to obtain equivariance to spherical rotations. In [28] and in [29], it was proposed to perform the convolutions in the spherical harmonic domain and then transform their result back to the spatial domain to pointwise apply the nonlinear activations. Since then, several modifications of this approach have been published, some of them proposing nonlinear activations in the harmonic domain [30], [31]. Another approach is to analyze the spherical signal as a graph where each point is connected to its neighbors [32], [33], [34]. By working this way, they avoid the need to work in the harmonic domain, but, in most cases, their kernels are restricted to being isotropic, i.e. to having circular symmetry.

In this work, we propose the use of a third approach: the icosahedral CNNs presented in [35]. These networks are only strictly equivariant to the 60 rotational symmetries of the icosahedron instead of the continuous space of spherical rotations, but they have a much more efficient implementation based on standard 2D convolutions. They have been proven to smoothly generalize to the continuous space of spherical rotations when these rotations are shown during the training of the model and their hexagonal kernels are not restricted to be isotropic thanks to saving the results of every possible orientation as separate channels.

Although the SRP-PHAT maps have a stronger equivariance to rotations when they are computed from the signals captured with a spherical array, it is worth mentioning that our proposed method could be used with any array geometry whose microphones are distributed in the three dimensions of the space; i.e., it is possible to compute the SRP-PHAT maps for any azimuth and elevation without ambiguity. To prove this, we use in our experiments an array geometry designed to be mounted in a NAO robot head [36], [37] which is neither spherical nor uniform.

In addition, in order to allow the network to take into account the temporal context and therefore keep a tracking of the source DOA even when it remains silent, we include 1D convolutions interleaved with the icosahedral convolutions.

Finally, for the output of the network, we use a regression approach, inferring a 3D vector pointing towards the DOA of the source but, to continue with our equivariant approach, we replace the fully connected layers that are typically employed after the convolutional layers in most CNNs by a new soft-argmax function. This function can be seen as a differentiable version of the argmax function and does not have any learnable parameters and allows us to interpret the output of the icosahedral CNN as a probability distribution.

To sum up, the main original contributions of this paper are i) the use of icosahedral CNNs over SRP-PHAT maps for DOA estimation and tracking of sound sources and ii) the use of a new soft-argmax function to perform the final regression after the CNN, replacing the traditional fully connected layers; in addition, we are making public the code needed to replicate the results presented in this paper[1], included what is, to the best of our knowledge, the first publicly available free and open-source implementation of the icosahedral CNNs[2]. In [38], the use of a CNN equivariant to spherical rotations for DOA estimation is also proposed, but they use an implementation based on a Graph Neural Network whose kernels are restricted to be isotropic [33], use a dense fully connected layer after the convolutions, and the accuracy of their estimations are not analyzed in deep.

The remainder of this paper is organized as follows. We first review the SRP-PHAT algorithm and the the icosahedral CNNs in sections II and III and present the soft-argmax function in section IV. Then, we present the details of our proposed model and its training in section V and in section VI we evaluate its accuracy against other state-of-the-art models using both simulated signals and actual recordings and discuss the obtained results. Finally, section VII summarizes the conclusions of the paper.

## II. SRP-PHAT POWER MAPS

The Steered Response Power (SRP) is a classical signal processing technique that allows us to obtain acoustic power maps by computing the energy received by several filter-and-sum beamformers steered at different directions [4], [5]. Although individually computing the output of every beamformer would be computationally intensive, we can compute the energy $P(\boldsymbol{\theta})$ coming from the direction $\boldsymbol{\theta}$ in terms of the Generalized Cross Correlations (GCCs) of the signals received by each microphone:

$$P(\boldsymbol{\theta}) = 2\pi \sum_{n=0}^{N-1} \sum_{m=0}^{N-1} R_{nm}(\Delta\tau_{nm}(\boldsymbol{\theta})), \quad (1)$$

where $N$ is the number of microphones, $\Delta\tau_{nm}(\boldsymbol{\theta})$ is the time difference of arrival between the $n$ and the $m$ microphone for a signal coming from $\boldsymbol{\theta}$, and $R_{nm}(\tau)$ is the GCC of the signals received at the $n$ and the $m$ microphones:

$$R_{nm}(\tau) = \frac{1}{2\pi} \int_{-\infty}^{+\infty} \Psi_{nm}(\omega) X_n(\omega) X_m^*(\omega) e^{j\omega\tau} d\omega, \quad (2)$$

where $*$ denotes the complex conjugate, $X_n(\omega)$ is the Fourier Transform of the signal received at the microphone $n$, and $\Psi_{nm}(\omega)$ is a weighting function. Due to its good performance

---

[1] https://github.com/DavidDiazGuerra/icoDOA
[2] https://github.com/DavidDiazGuerra/icoCNN

in reverberant scenarios [39], the Phase Transform (PHAT) is typically used:

$$\Psi_{nm}(\omega) = \frac{1}{|X_n(\omega)X_m^*(\omega)|} \quad (3)$$

It is worth saying that $\boldsymbol{\theta}$ can represent an angle, two spherical coordinates, or even a point in 3D Cartesian coordinates depending on the geometry of the array. In this paper, since we focus on compact 3D arrays, we work with spherical coordinates and, in this case, we can create an icosahedral power map just by computing (1) for the elevation and azimuth angles of the desired icosahedral grid as done in Fig. 1.

The position of the maximum of the SRP-PHAT maps have traditionally been used as the DOA estimation; however, as we can see in Fig. 1, in scenarios with a high reverberation time, the map maximum may be too wide and lead to inaccurate estimates or even to erroneous estimates when spurious maxima appear. To fix this issue, we propose the use of a CNN over these maps but, since they are equivariant to spherical rotations instead of Cartesian translations, we use icosahedral CNNs to better approximate the rotational symmetry of the problem.

## III. ICOSAHEDRAL CNNs

Several techniques have been recently proposed to extend the translation equivariance of conventional CNNs to spherical rotations, most of them based on the Spherical Harmonics domain. Although they are only equivariant to the 60 icosahedral rotations instead of the continuous space of spherical rotations SO(3), in this work we use the icosahedral CNNs proposed in [35] due to its efficient implementation based on conventional bi-dimensional CNNs. The icosahedron is the platonic solid with the highest number of faces and it allows us to approximate a sphere with lower error than other geometric shapes with a lower number of faces while being able to define a convolution operation on a hexagonal grid without needing any kind of interpolation and implement it using a conventional 2D convolution. The details of this implementation and some experiments proving its good performance approximating spherical signals (and its rotations) can be found in [35], but we summarize some of the main ideas in this section.

The icosahedral grid used to sample the spherical signals is built recursively starting from the vertices of the icosahedron (which we define having a vertex in the south and another in the north pole as shown in Fig. 1a) and then subdividing each triangular face into 4 smaller triangles by introducing a new point in the center of each edge. Repeating this process $r$ times, we obtain a grid with $5 \cdot 2^{2r+1} + 2$ points. As we can see in Figs. 1a and 1c, apart from the vertices of the icosahedron, which only have 5, every one of these inner sampling points has 6 neighbors, so we can see them as hexagonal pixels. In [35], due to their pentagonal shape, it is proposed to keep the corners values to 0 to preserve the equivariance of the model but, in order to avoid artifacts around them after applying the convolutions, we replace that 0 by the average value of their 5 neighbors (which also preserves the equivariance of the model).

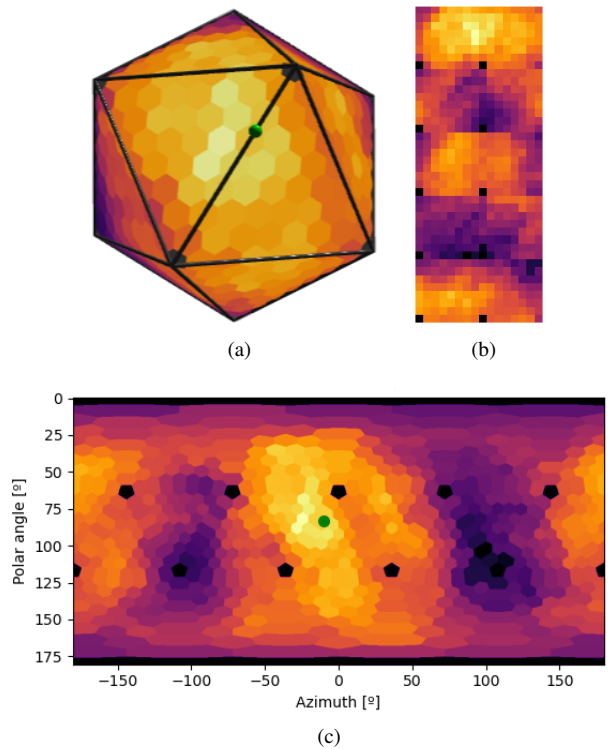

Fig. 1. Example of an icosahedral SRP-PHAT power map with resolution $r = 3$ in a high reverberation low noise scenario: $T_{60}=1.0\,\text{s}$ and SNR=30 dB (a), the 2D representation employed for the convolution implementation (b) and a spherical projection for visualization purposes (c). The green sphere/circle indicates the actual DOA of the sound source.

Due to the hexagonal shape of the pixels defined this way, the icosahedral CNNs use hexagonal kernels which can be stored in $3 \times 3$ bi-dimensional kernels. If we want the model to be equivariant to the icosahedral rotations without restricting the kernels to be isotropic, we have to consider their 6 possible rotations so, for each kernel in the convolution, we have to work with 6 channels instead of with just one as it is done in conventional convolutions. Finally, in order to implement the icosahedral convolutions using standard 2D convolutions, a projection of the whole icosahedral grid into a $5 \cdot 2^r \times 2^{r+1}$ rectangular grid (Fig. 1b) is presented in [35] along with a way to circularly pad it into a $5 \cdot (2^r + 2) \times 2^{r+1} + 2$ extended grid to preserve the equivariance of the model.

To sum up, we can apply a convolution with $C$ kernels over an icosahedral grid of resolution $r$ using a conventional 2D convolutional layer with $3 \times 3$ kernels and $6C$ channels over a $5 \cdot (2^r + 2) \times 2^{r+1} + 2$ image.

In [35], it is not detailed how the pooling layers are implemented. In our implementation, an icosahedral pooling layer reduces an icosahedral grid of resolution $r$ to a new one of resolution $r - 1$ where each new hexagonal pixel is computed as the average of the pixel which has its same center and its 6 neighbors. This can be seen as the icosahedral equivalent of a 2D pooling with kernel size $3 \times 3$ and stride $2 \times 2$.

## IV. Soft-argmax regression

In [27], in order to perform the regression of the source coordinates after the convolutional layers, we flattened the activation maps of the last layer and fed it to several fully connected layers (actually, we used 1D convolutions in order to allow the model to also take into account the previous frames). Even when employing pooling layers along with the convolutional layers to reduce the number of activations that reach the fully connected layers, this approach has a high computational cost and highly increases the number of trainable parameters of the model, which increases its memory consumption and its risk of overfitting.

In this proposal, we replace those fully connected layers by a new soft-argmax function, where we use a soft-max layer to ensure that the sum of the whole activation map is 1.0 and then we just sum the results of multiplying each hexagonal pixel by the coordinates that they represent in the icosahedral grid:

$$\text{soft-max}\left(P\left(\boldsymbol{x}\right)\right) = \frac{e^{P(\boldsymbol{x})-\max(P(\boldsymbol{x}))}}{\sum_{\boldsymbol{x}\in\mathcal{X}} e^{P(\boldsymbol{x})-\max(P(\boldsymbol{x}))}} \quad (4)$$

$$\text{soft-argmax}\left(P\left(\boldsymbol{x}\right)\right) = \sum_{\boldsymbol{x}\in\mathcal{X}} \boldsymbol{x}\,\text{soft-max}\left(P\left(\boldsymbol{x}\right)\right) \quad (5)$$

where $P(\boldsymbol{x})$ is the output of the last convolutional layer of the model and $\boldsymbol{x} \in \mathcal{X}$ are the coordinates of the points of the icosahedral grid $\mathcal{X}$ where it is sampled. The subtractions of $\max(P(\boldsymbol{x}))$ inside the exponential functions are done for numerical stability reasons without affecting the analytical result.

This way, the output of the icosahedral convolutions, after being normalized with the soft-max function, can be seen as the probability distribution of the coordinates of the source and the output of the soft-argmax function as its expected value. Another advantage of this approach, apart from the model interpretability and the reduction in the computational and memory costs, is that we avoid introducing any non-equivariant layer to the model.

Although we could directly estimate the spherical coordinates of the source defining $\boldsymbol{x}$ in spherical coordinates, we estimate the 3D coordinates of the unitary vector pointing in the direction of the source by defining $\boldsymbol{x}$ in 3D Cartesian coordinates. It has been proven that this brings better results than directly inferring the elevation and azimuth angles [22], [23] and, in addition, continuing with the interpretation of the output of the CNN as a probability distribution function, minimizing the Mean Square Error (MSE) of this vector does not only imply reducing the distance between its expected value and the source position but also reduces its variance, since a completely unitary vector would only be possible if only one pixel is activated. Therefore, we can see the norm of the output vector as a measurement of the confidence in the DOA estimation, being closer to 1 when the confidence is higher.

In Fig. 2 we can see an example of the output of the last convolutional layer of the trained model after being normalized with the soft-max function and how it represents a probability

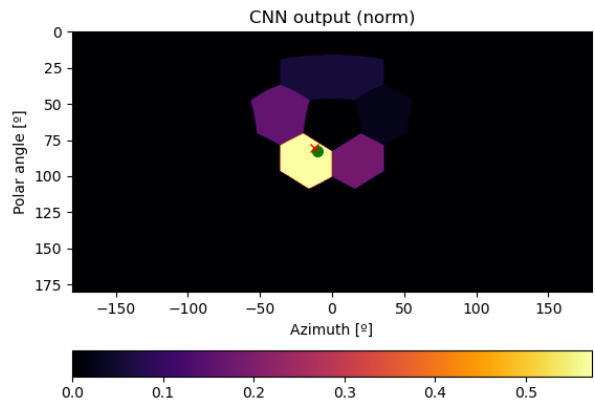

Fig. 2. The output of the last convolutional layer after passing through the soft-max function, corresponding to the input maps shown in Fig 1, and the result of the soft-argmax function converted into spherical coordinates (red cross). The green circle indicates the actual DOA of the source.

distribution whose expected value accurately estimates the actual DOA of the sound.

## V. Proposed technique

### A. Model architecture

As shown in Fig. 3, we combine icosahedral convolutions with one-dimensional convolutions operating in the time dimension in order to take the temporal context into account when performing the DOA estimation while being equivariant to both icosahedral rotations and temporal translations. Each convolutional unit is composed of an icosahedral and a temporal convolution followed by layer normalization [40] and Rectified Linear Unit activation. The temporal convolution has a kernel of size 5 operating causally, i.e. its Receptive Field (RF) only includes past maps, and both the icosahedral and the temporal convolution have 32 kernels. To preserve the equivariance of the model, the 6 kernel-orientation channels of every icosahedral kernel are seen as 6 independent signals by the temporal convolutions. We did not observe significant improvements using more than 32 kernels in the convolutions, but this might be optimized using a different number of kernels in every layer.

In [40], it was hypothesized that layer normalization did not provide relevant improvements in convolutional neural networks since the hidden units close to the boundaries of the images did not follow the same distribution as the rest of the hidden units. However, our inputs do not have boundaries and we found that, adding layer normalization to our model, it converged faster and more robustly during the training. To keep the model equivariant to the icosahedral rotations, we have implemented a layer normalization that normalizes the inputs along the 32 channels and its 6 kernel orientations but the scale weights are tied for the 6 kernel orientations in the affine transformation.

We concatenate two of these convolutional units to an icosahedral pooling to build a down-sampling unit and stack as many of these down-sampling units as needed to get an $r = 1$ icosahedral map (in the case of using maps with $r = 1$ as input we do not use any down-sampling unit). When we have a minimal-size icosahedral map, we concatenate 5 convolutional units (the last one with only an output channel for the 1D convolution and without layer normalization and the ReLU activation). This way, we ensure that the receptive field of all the output cells of the icosahedral CNN includes all the cells of the input map independently of its resolution. Finally, we use a max-pooling layer over the 6 kernel-orientation channels and feed the resulting icosahedral maps to the soft-argmax layer explained in section IV.

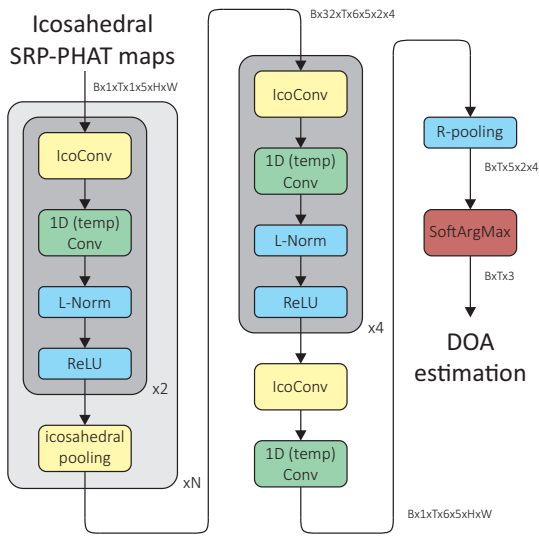

Fig. 3. Architecture of the proposed model. $B$ is the batch size, $T$ is the number of temporal frames of the trajectory, $H = 2^r$ and $W = 2^{r+1}$ are the height and the width of the projections of the icosahedral grid, and $N = r-1$ is the number of down-sampling units used for that input resolution.

TABLE I
MODELS EMPLOYED FOR THE EVALUATION

| Model | Input | SRP-PHAT computations | | Trainable parameters | Temporal RF |
|---|---|---|---|---|---|
| IcoCNN | Icosahedral SRP-PHAT maps | r=1 | 30 | 193 505 | 4.10 s |
| | | r=2 | 150 | 290 017 | 5.63 s |
| | | r=3 | 630 | 386 529 | 7.17 s |
| | | r=4 | 2550 | 483 041 | 8.70 s |
| Cross3D | Equiangular SRP-PHAT maps | 4x8 | 18 | 526 372 | 5.63 s |
| | | 8x16 | 98 | 946 340 | 6.40 s |
| | | 16x32 | 450 | 1 693 988 | 7.17 s |
| | | 32x64 | 1922 | 5 626 148 | |
| | | 64x128 | 7938 | 21 354 788 | |
| 1D CNN | GCCs | | 0 | 11 282 436 | 7.17 s |

### B. Training

For the training dataset, we followed the technique described in [27] to generate random source trajectories and simulated them using a GPU implementation [41] of the Image Source Method [42] at a sample rate of 16 kHz using utterances from the LibriSpeech train-clean-100 dataset [43] as source signals. We simulated the 12-microphones array included in the LOCATA dataset [36], [37] designed to be mounted over a NAO robot head, which has a minimum and maximum inter-microphone distances of 1.3 cm and 12.1 cm respectively.

The sizes of the simulated rooms were randomly sampled from the range $3\,\text{m} \times 3\,\text{m} \times 2.5\,\text{m}$ to $10\,\text{m} \times 8\,\text{m} \times 6\,\text{m}$ and the absorption coefficients were randomly adjusted to obtain an uniform distribution of reverberation times form $T_{60} = 0.2\,\text{s}$ to $1.3\,\text{s}$. The array was randomly placed inside the room, being restricted to have a minimum separation from the walls of a 10% of the room size in each dimension and be in the lower half of the room for the vertical axis. For the source trajectories, we randomly chose two points inside the room to be the starting and ending points of the trajectory and, to the straight line that connect them, we added a sinusoidal function in each axis with random frequencies and amplitudes ensuring that no more than 2 oscillations were done in each axis during the trajectory and that the source never left the room boundaries.

As done in [27], we computed the SRP-PHAT maps using frames of length $K = 4096$ samples (i.e. 256 ms) with a hop size of $3K/4$. Also as described in [27], we normalized the maps subtracting their mean and dividing them between their maximum and used the Voice Activity Detector of the WebRTC project [44] to turn to 0 the maps corresponding to silent frames.

Using this approach, we have an infinite-size dataset, but we define an epoch as 585 random trajectories of 20 s, each one with an utterance randomly taken from one of the 585 chapters present in the LibriSpeech train-clean-100 subset. We used Pytorch [45] to train the model using the Adam algorithm [46] over 50 epochs. Similar to the curriculum learning [47] strategy employed in [27], we keep fixed the SNR of the simulations to 30 dB during the first 25 epochs and then we employed uniformly distributed random values from 5 dB to 30 dB in the following epochs.

## VI. EVALUATION

### A. Baseline techniques

Since they were the techniques that proved to be the most robust against reverberation in [27], we compared the proposed technique with Cross3D, a 3D CNN working over rectangular equiangular SRP-PHAT maps followed by 1D temporal convolutions acting as fully connected layers, and against the 1D CNN operating over the GCC coefficients also described in [27].

As we can see in Table I, the proposed model has a much lower number of trainable parameters than the other models. This is because the final regression is performed with the soft-argmax function (without trainable parameters) instead of with fully connected layers.

### B. Simulated dataset

In order to analyze the performance of the proposed model under different acoustic conditions, we first evaluated our model using synthetic signals simulated following the same procedure employed for the training dataset. We found that some utterances included a short period of silence at the beginning which artificially biased the results, so we have not taken into account the localization error of the first 5 frames, i.e. the first second, of each trajectory even if they were used to train the models. All the root mean localization errors described in this paper include the frames where the source was silent, but the models are able to continue estimating its position thanks to the temporal context included in their Receptive Fields (RF) through the temporal convolutions.

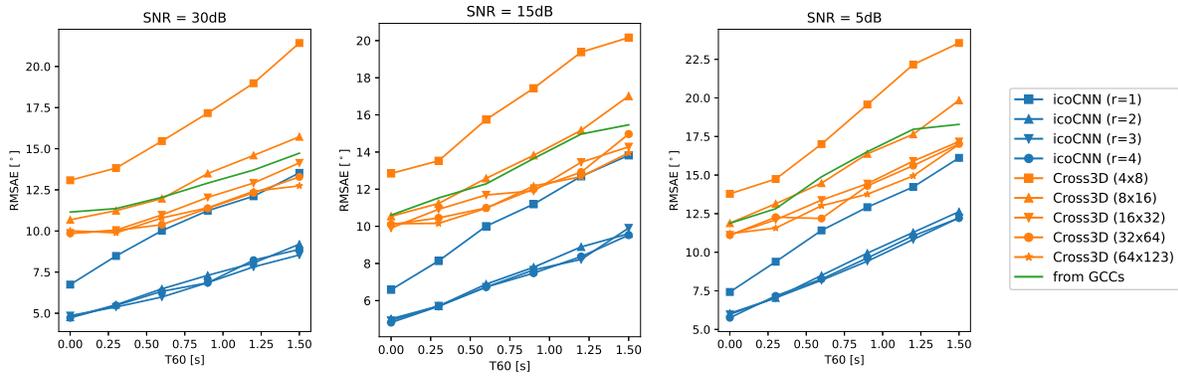

Fig. 4. Localization Root Mean Squared Angular Error (RMSAE) under several simulated conditions for the proposed and the baseline techniques.

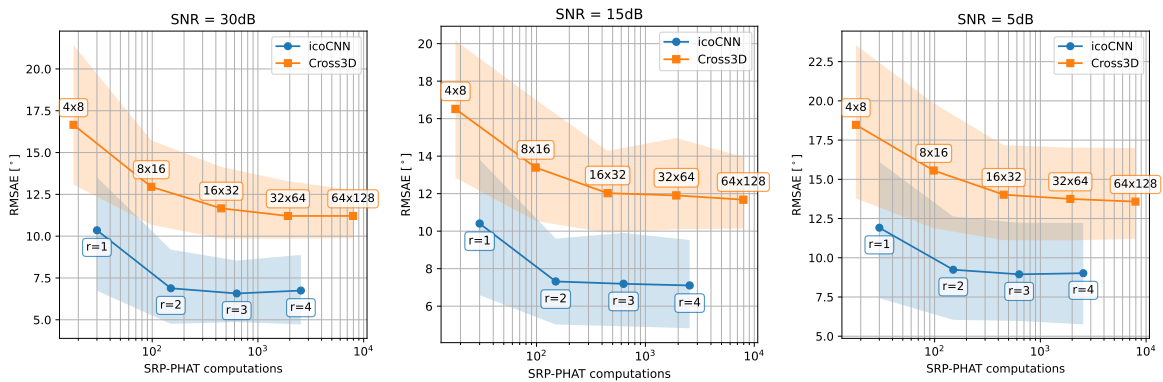

Fig. 5. Localization Root Mean Squared Angular Error (RMSAE) vs the number of computations of equation (1) needed to compute each input map. The semitransparent area indicates the whole reverberation interval from $T_{60} = 0.2$ s to $1.3$ s and the solid line indicates its average value.

As we can see in Fig. 4, the proposed models clearly outperform the baselines even using maps of lower resolution and having far less trainable parameters. To make this even clearer, Fig. 5 plots the localization error represented as a function of the number of computations of the SRP-PHAT functional (1), needed to compute the input maps of each model. It is worth noting that the reverberation times $T_{60} = 0.0$ s and $T_{60} = 1.5$ s are out of the range of reverberation times used during the training, but the model generalizes well to them continuing with the same tendency shown with the rest of reverberation times.

We can also see how using icosahedral maps with a resolution higher than $r = 2$ does not seem to improve the accuracy of the DOA estimations. Considering that those models had a higher number of down-sampling units, and therefore more trainable parameters and longer temporal receptive fields, we can conclude that the maps with $r = 2$ already contain all the information useful for tracking. This limit is probably determined by the size of the array we are using to compute the maps and that limits their spatial bandwidth; using bigger arrays would probably allow us to obtain even better results from higher resolution maps.

In Fig. 6a, we can see an example of a random trajectory in a scenario with high reverberation time and low noise using maps with resolution $r = 2$. We can see that the maximums of the SRP-PHAT maps are in spurious positions in many frames and, even in those where they are in the grid position closest to the ground truth, they are quite far due to its low resolution. However, we can see that the estimation of the proposed model stays always closer to the actual DOA of the source even during the silent frames since the icosahedral convolutions allow the model to analyze the whole maps instead of just taking only the information of the position of their absolute maximums and thanks to the temporal context provided by the temporal convolutions.

In Fig. 2 we could see the $r = 1$ probability distribution inferred by the model for the $r = 3$ maps depicted in Fig. 1. We can see how the model is able to accurately adjust the probability assigned to every hexagonal pixel around the pentagonal vertex so the result of the soft-argmax function is precisely displaced from the center of the closest hexagonal pixel to the actual DOA of the simulated source. It is worth saying that, even if the vertex values were replaced by the average value of their neighbors during the convolutional layers, they are turned to 0 before the soft-argmax function.

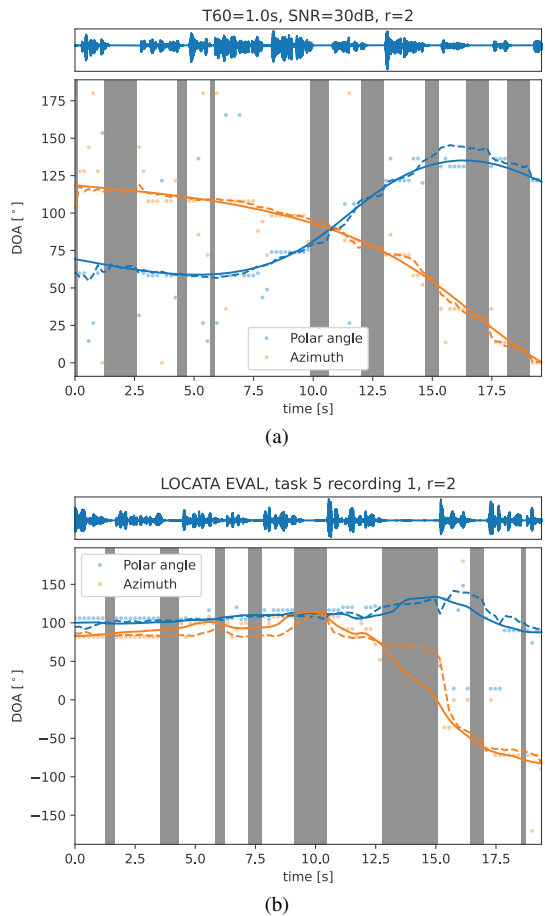

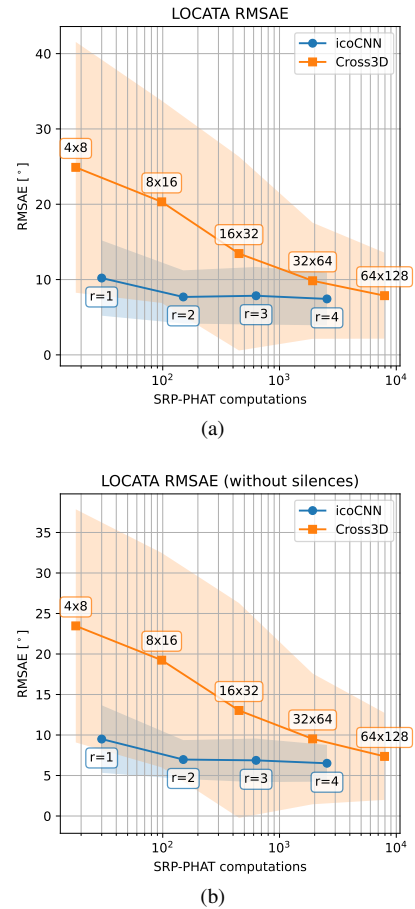

Fig. 6. DOA estimation for a random trajectory simulated with $T_{60}$=1.0 s and SNR=30 dB (a) and for the third recording of the task 5 of the LOCATA dataset (b). The solid lines represent the ground-truth DOA and the dashed lines the estimation performed by the proposed method using input maps of resolution $r = 2$. The dots indicate the position of the maximum of those maps.

Fig. 7. Average localization Root Mean Squared Angular Error (RMSAE) in the LOCATA dataset vs the number of computations of equation (1) needed to compute each input map, along the whole signal (a) and without taking into account the silent frames (b). The semitransparent area indicates the average $\pm$ the standard deviation.

### C. LOCATA dataset

To confirm that the models trained with simulated signals are general enough to work with signals recorded in real environments, we have tested them using the recordings of tasks 1, 3, and 5 (i.e. the tasks with only one source) of the evaluation partition of the LOCATA dataset [36]. Table II and Fig. 7 show the average results for every task and for the whole evaluation partition and we have included the RMSAE of every recording in the supplementary materials of this paper.

As we can see in Fig. 6b, some of the recordings of the LOCATA dataset include long periods of silence when the sound source was moving. During these silent periods, the models obviously cannot keep tracking the movements and if they constitute a high percentage of a recording, its RMSAE will be strongly biased by them; therefore, Table II also include the RMSAEs without taking into account these silent frames. In addition, it is worth mentioning that the size of the dataset is quite low, with the 23 recordings used for this test adding less than 7 minutes after having removed the initial silences, which makes it very sensitive to any anomalous circumstance that could appear even if it is only present for a short period of time.

We can see that the difference between the proposed model and the model using 3D CNNs over equiangular maps proposed in [27] is not as great in this dataset as in the synthetic one, though the model using icosahedral convolutions still clearly outperforms the baseline model, especially when using low resolution power maps. This could be due to the differences between the signals simulated with the ISM method used to train the models and the signals recorded in a real room used for this test. It seems that we might be approaching the accuracy limit imposed by this dataset difference, so the accuracy with real recordings can not improve even when we improve the models. In recent years, several domain adaptation techniques have been proposed to improve the accuracy of models trained with simulated signals [48], [49], [50], [51] and it would be interesting to conduct further studies along these lines.

In any case, as can be seen in Fig. 7, the proposed model working with maps of resolution $r = 2$ still has in the LOCATA dataset an accuracy comparable with Cross3D using maps of much higher resolution, reducing in almost two orders of magnitude both the number of SRP-PHAT computations and the number of trainable parameters, which might be crucial in applications where the sound source localization must be done in low-cost devices in real time. We can also see how the proposed model provides far more consistent results than the baseline (whose results have a much higher variance), which also suggest a better generalization from the simulated dataset used for training to the actual recordings of the LOCATA dataset.

TABLE II
MEAN RMSAE [°] OF THE DOA ESTIMATED FOR EVERY TASK OF THE EVALUATION PARTITION OF THE LOCATA DATASET WITH THE ICOSAHEDRAL CNNS USING SEVERAL MAP RESOLUTIONS AND THE BASELINE TRACKING METHODS. THE SECOND (GRAY) NUMBERS INDICATE THE RMSAE WITHOUT TAKING INTO ACCOUNT THE FRAMES WHEN THE SOUND SOURCE WAS SILENT.

| Model: | IcoCNN | | | | Cross3D | | | | | 1D CNN |
|---|---|---|---|---|---|---|---|---|---|---|
| Input: | Icosahedral SRP-PHAT maps | | | | Equiangular SRP-PHAT maps | | | | | GCCs |
|  | r=1 | r=2 | r=3 | r=4 | 4x8 | 8x16 | 16x32 | 32x64 | 64x128 |  |
| Task 1 | 8.04 / 8.34 | 5.88 / 6.08 | 5.57 / 5.78 | 5.25 / 5.23 | 28.62 / 27.09 | 22.47 / 22.16 | 13.89 / 14.32 | 9.84 / 8.51 | 5.28 / 5.48 | 12.54 / 12.80 |
| Task 3 | 10.53 / 8.97 | 8.94 / 7.29 | 10.07 / 6.93 | 9.57 / 7.87 | 18.51 / 17.98 | 17.56 / 17.50 | 12.76 / 12.16 | 11.18 / 10.33 | 9.92 / 8.86 | 12.09 / 11.62 |
| Task 5 | 15.48 / 13.03 | 11.13 / 8.87 | 11.54 / 9.66 | 10.95 / 8.49 | 19.13 / 15.82 | 17.49 / 13.38 | 13.03 / 10.75 | 12.20 / 10.63 | 12.49 / 10.73 | 16.47 / 13.31 |
| Average | 10.20 / 9.50 | 7.69 / 6.97 | 7.85 / 6.87 | 7.43 / 6.51 | 24.90 / 23.47 | 20.32 / 19.24 | 13.45 / 13.03 | 9.84 / 9.52 | 7.86 / 7.36 | 13.30 / 12.65 |
| Median | 9.91 / 9.94 | 7.17 / 6.79 | 6.66 / 6.53 | 6.60 / 6.45 | 18.51 / 18.25 | 15.32 / 14.43 | 9.90 / 7.65 | 7.58 / 7.09 | 5.97 / 5.74 | 12.72 / 12.00 |
| Standard deviation | 5.00 / 4.17 | 3.53 / 2.40 | 3.82 / 2.69 | 3.51 / 2.26 | 15.65 / 14.39 | 13.43 / 13.24 | 12.87 / 13.26 | 7.71 / 8.07 | 5.71 / 5.37 | 5.67 / 5.44 |

Finally, Fig. 8 shows an example of an input SRP-PHAT map extracted from the first recording of the fifth task of the evaluation partition of the LOCATA dataset and its corresponding model output. As in Fig. 2, we can see how the energy of output of the CNN is distributed in a way that, after normalized with the soft-max function and interpreted as a probability distribution, its expected value approach the actual DOA of the sound.

## VII. CONCLUSIONS

We have presented a new model for direction of arrival estimation of sound sources that is completely equivariant to time shifts and to the 60 rotational symmetries of the icosahedron, which is a good approximation of the continuous space of spherical rotations. This model can be implemented using conventional 2D convolutional layers and has a low number of trainable parameters thanks to replacing the fully connected layers that are typically employed after the convolutional layers with a differentiable version of the argmax function, which, in addition, allows us to interpret the output of the convolutional layers as a probability distribution whose expected value is the DOA estimation.

This new model outperforms the state of the art in terms of accuracy and efficiency, especially in highly reverberant scenarios, even when using SRP-PHAT maps with only 150 points. Testing it with simulated signals, we prove that it can estimate the DOA of a sound source in extremely adverse scenarios with a Root Mean Square Angular Error (RMSAE) lower than 10° using a really compact array and, even in a challenging dataset with real recordings, it maintained its average RMSAE under 10°. However, further studies using domain adaptation techniques would be interesting to improve even more its accuracy with real world signals.

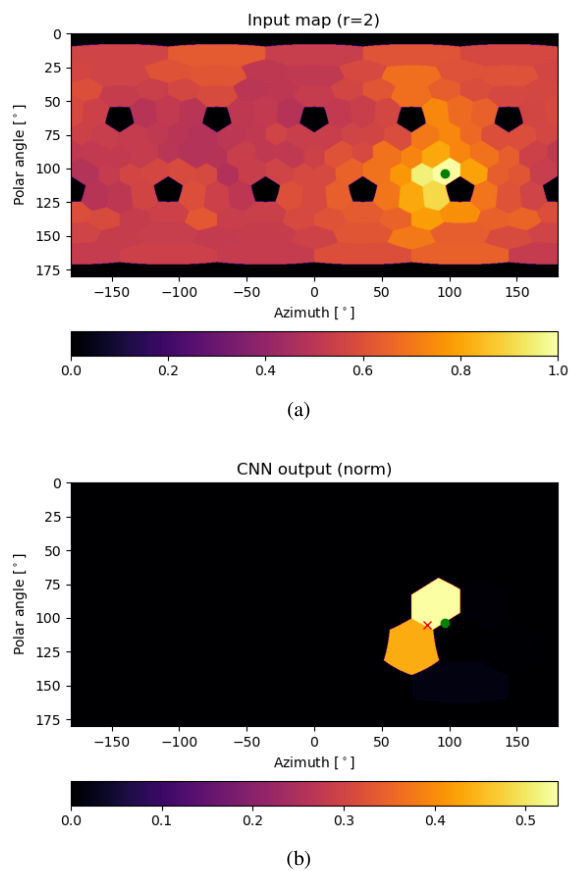

Fig. 8. Example of an input map from the 1st recording of the 5th task of the LOCATA evaluation dataset (a) and the output of the last convolutional layer after passing through the soft-max function and the result of the soft-argmax function converted into spherical coordinates (red cross) (b). The green circle indicates the actual DOA of the source.

Finally, it is worth mentioning that, even if we have focused on using SRP-PHAT maps as inputs for the icosahedral CNNs, any other spatial pseudo-spectrum could also be used (such as those computed using the MUSIC algorithm), so further studies comparing different input features and even combining them as different input channels would also be interesting.